\input amstex
\input amsppt.sty
\input epsf
\input crossref
\nologo
\advance\hoffset by-5truemm
\pagewidth{170truemm}
\pageheight{247truemm}
\input epsf
\raggedbottom
\TagsOnRight

\def\Real{{\Bbb R}}
\def\sobnorm#1#2#3{{\Vert#3\Vert}^{#1}_{#2}}
\def\sobnormsq#1#2#3{\bigl(\sobnorm{#1}{#2}{#3})^2}
\let\leq=\leqslant

\let\epsilon=\varepsilon
\def\cd{\Cal{D}}
\def\ip#1#2{\langle#1,\,#2\rangle}
\def\flux#1#2{F(#1)}
\def\energy#1#2{E(#1)}

\topmatter
\title
Generalised hyperbolicity: hypersurface singularities
\endtitle
\author
J.~A.~Vickers and J.~P.~Wilson
\endauthor
\address
Faculty of Mathematical Studies, University of Southampton, Highfield,
Southampton SO17 1BJ, UK.
\endaddress
\email
J.A.Vickers@maths.soton.ac.uk, J.P.Wilson@maths.soton.ac.uk
\endemail
\abstract
Sufficient conditions for the well-posedness of the initial value problem
for the scalar wave equation are obtained in space-times with hypersurface
singularities
\endabstract
\endtopmatter

\document
\crossreferencing

\head 1. Introduction \endhead

A desirable property of any space-time used to model a physically plausible
scenario is that the evolution of the Einstein's equations is well posed;
that is the initial value problem admits a unique solution.  Space-times
whose metrics are at least $C^{2-}$, which guarantees the existence of
unique geodesics, fall within the context of the Cosmic Censorship
Hypothesis of Penrose~(1979). This hypothesis states that that the
space-time will be globally hyperbolic, i.e. strong causality is satisfied
and $J^+(p)\cap J^-(q)$ is compact $\forall p,\,q\in M$, and hence the
evolution of Einstein's equations is well defined.

There are however a number of space-times which violate cosmic censorship
in this sense, but for which there may be a well posed initial value
theorem for test fields. This suggests that a weaker notion of
hyperbolicity should be defined (Clarke,~1996).  Space times which may fall
into this category include those with weak singularities such as as thin
cosmic strings (Vickers,~1987), impulsive gravitational waves
(Penrose,~1972) and dust caustic space-times (Clarke and O'Donnell,~1992),
all of which model physically plausible scenarios. Such space-times
typically admit a locally bounded metric whose differentiability level is
lower than $C^{2-}$ and a curvature tensor that is well defined as a
distribution, often with its support on a proper submanifold.

A concept of hyperbolicity for such space-times was proposed by
Clarke~(1998). This was based on the extent to which singularities
disrupted the local evolution of the initial value problem for the scalar
wave equation.
$$ \aligned
   \square \phi & =f \\
   \phi_{|S} &= \phi_0 \\
   n^a {\phi_{,a}}_{|S} &= \phi_1
   \endaligned
$$
Clarke reformulated the initial value problem, on an open region $\Omega$
with a compact closure admitting a space-like hypersurface $S$, which
partitions $\Omega$ into two disjoint sets $\Omega^+$ and $\Omega^-$, in a
distributional form, obtained by multiplying by a test field $\omega$ and
integrating by parts once to give
$$ \aligned
   &\int_{\Omega^+}\!\!\! \phi_{,a} \omega_{,b} g^{ab} \,(-g)^{1/2}\,d^4x
   = -\int_{\Omega^+}\!\!\!f \omega\,  (-g)^{1/2} \,d^4x
   - \int_{S} \phi_1 \omega \, dS \qquad \forall\omega\in\cd(\Omega) \\ 
   \noalign{\medskip}
   &\phi_{|S}=\phi_0
   \endaligned
$$
and then defined a point $p\in M$ as being {\it $\square$-regular} if it
admitted such a neighbourhood $\Omega$ for which the above equation had a
unique solution for each set of Cauchy data $(\phi_0,\phi_1)\in
H^1(S)\times H^0(S)$.  A space-time which was $\square$-regular everywhere
could then be said to be {\it $\square$-globally hyperbolic}.  It was shown
if a space-time satisfied the following curve integrability conditions at a
given point $p\in M$, then that point was $\square$-regular:
\roster
\item $g_{ab}$ and $g^{ab}$ are continuous
\item $g_{ab}$ is $C^1$ on $M-J^+(p)$
\item $g_{ab,c}$ exists as a distribution and is locally square integrable.
\item The distributional Riemann tensor components $R^{a}{}_{bcd}$ may be
interpreted as locally integrable functions
\item There exists a non empty open set $C\subset\Real^4$ and functions
$M,N:\Real^+\to\Real^+$, with $M(\epsilon),\,N(\epsilon)\to0$ as $\epsilon\to0$, such that if
$\gamma:[0,1]\to M$ is a curve with $\dot\gamma\in C$ then $\gamma$ is
future time-like and
$$ \aligned
   &\int^\epsilon_0 \,\bigl| \Gamma^a_{bc}(\gamma(s)) \bigr|^2\,ds <
   M(\epsilon) \\ 
   &\int^\epsilon_0 \,\bigl| R^{a}{}_{bcd}(\gamma(s))\bigr| \,ds < N(\epsilon)
   \endaligned
$$
\endroster
The proof involved the construction of a congruence of time-like geodesics,
whose tangent admitted an essentially bounded weak derivative, and a
suitable energy inequality from which uniqueness and existence could be
deduced. In particular it was shown that these results were applicable to
the dust caustic space-times.

There is an important class of space-times with weak singularities that do
not satisfy these curve integrability conditions; space-times whose
curvature has support on a proper submanifold. This class includes both
impulsive gravitational waves (Co-dimension 1) and thin cosmic strings
(Co-dimension 2).

This paper will consider the question of hyperbolicity in space-times with
a singular hypersurface. For hyperbolicity in conical space-times see
Wilson (2000) and also Vickers and Wilson (2000).

\head 2. Existence and uniqueness \endhead

In order to prove hyperbolicity, we must show that in each open region
$\Omega$ with a compact boundary $\partial \Omega$, that a unique solution
$\phi\in H^1(\Omega)$ exists to the initial value problem;
$$ \aligned
   \square \phi & =f \\
   \phi_{|S} &= \phi_0 \\
   n^a {\phi_{,a}}_{|S} &= \phi_1
   \endaligned
   \tag\tagnum\xlabel{ivp}
$$
where $f\in H^1(\Omega)$ and the initial data $\phi_0$ and $\phi_1$ lie in
suitable function spaces. We shall assume that $(\phi_0,\phi_1)\in
H^1(S)\times H^0(S)$ at least; this is what was required for hyperbolicity
in curve integrable space-times (Clarke, 1998). However we shall see that,
in the given circumstances, that one requires some what stronger conditions
on $(\phi_0,\phi_1)$.

In an open region $\Omega$ with a closed boundary, we shall assume that we
have the following scenario
\roster
\item A local coordinate system $(t,x^\alpha)$ exists for $\Omega$ such
that the initial surface $S$ is described by $t=0$ and $g_{ab}$ and
$g^{ab}$ are essentially bounded (i.e. bounded almost everywhere)
\item There exists a hypersurface $\Lambda$ such that $g_{ab}$ is $C^{2-}$,
on $\Omega-\Lambda$. This characterises $\Lambda$ as a
singular hypersurface in a regular space-time.
\item $g_{ab}$ is $C^{1-}$ on $\Omega$
\item A congruence of time-like geodesics exists whose tangent field $T^a$
has an essentially bounded covariant derivative. Without loss of generality
we shall assume that $t$ is proper time along each of these geodesics;
this means that $T^a=\delta^a_t$ and $g_{tt}=-1$.
\endroster

We shall also assume without loss of generality that $\Omega$ is generated
by a foliation of space-like hypersurfaces $(S_\tau)_{a<\tau<b}$ (with
$a<0<b$) having a common boundary, with $S_0$ coinciding with the initial
surface $S$ (see figure 1).

\topinsert
\null
\bigskip
\bigskip
\line{\hfill\epsfbox{hyperpic.1}\hfill\hfill\hfill}
\bigskip
\bigskip
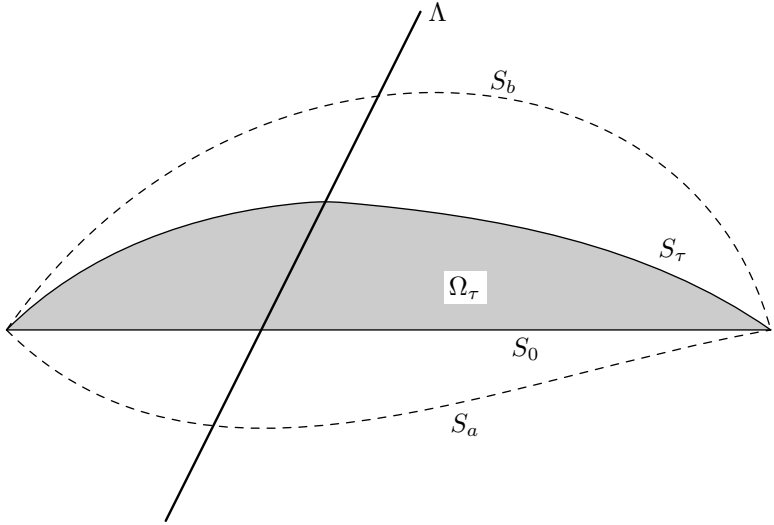
\centerline{Figure 1. The foliation of region $\Omega$}
\endinsert

\subhead 2.1 The energy inequality \endsubhead

We shall follow the approach of Hawking and Ellis~(1973), Clarke~(1998) and
Wilson~(2000) in discussing existence and uniqueness of solutions
to~\eqref{ivp} in terms of an energy inequality. The energy inequality
gives a bound to the solution $\phi$ at a given time in terms of the
initial data $(\phi_0,\phi_1)$ and the source term $f$.

We define the following energy integral
$$ \energy\tau\phi = \int_{S_\tau}\!\!\! S^{ab} T_a n_b \,dS $$
where $n^a$ is the future pointing normal vector to
the surface $S_\tau$ and
$$ S^{ab} = (g^{ac}g^{bd} - \tfrac12 g^{ab}g^{cd})\phi_{,c} \phi_{,d} -
   \tfrac12 g^{ab} \phi^2 $$
It will turn out that estimating this energy integral easier than working
directly with the classical Sobolev norm
$\sobnorm{1}{S_\tau}{\phi}$. However $\energy\tau\phi$ and
$\sobnormsq{1}{S_\tau}{\phi}$ are equivalent in the sense that
there exist positive constants $B_1$ and $B_2$ such that for
$a \leq\tau\leq b$
$$ B_1 \sobnormsq{1}{S_\tau}{\phi} \leq \energy\tau\phi \leq B_2
   \sobnormsq{1}{S_\tau}{\phi} \tag\tagnum\xlabel{sandwich} $$
See Wilson~(2000) for a proof of this result.

\proclaim{Lemma \procnum}
Solutions $\phi\in H^1(\Omega^+)$ of the initial value problem~\eqref{ivp}
satisfy an energy inequality 
\endproclaim

\demo{Proof}
In order to obtain an energy inequality we must apply Stokes' theorem to
the vector $S^{ab}T_b$ on the region
$$ \Omega_\tau= \bigcup_{0<\tau'<\tau} S_{\tau'} $$
however caution must be exercised because of the possible lack of
differentiability of $g_{ab}$ and $\phi$ in a neighbourhood of $\Lambda$.
We shall therefore apply Stokes' theorem to a sequence of regions
$\Omega^\epsilon_\tau$ which converges to $\Omega_\tau$. Without loss of
generality we shall illustrate the procedure for the case where $\Lambda$
intersects each $S_\tau$ exactly once, partitioning $\Omega$ exactly into
two regions (Figure 2).

\topinsert
\null
\bigskip
\bigskip
\line{\hfill\epsfbox{hyperpic.2}\hfill\hfill\hfill}
\bigskip
\bigskip
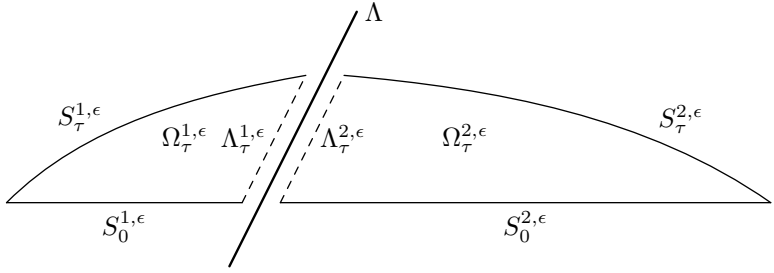
\centerline{Figure 2. Applying Stokes' theorem to the region $\Omega$}
\endinsert

On applying Stokes' theorem to both regions we have
$$ \aligned
    \int_{\Omega_\tau^{1,\epsilon}\cup\Omega_\tau^{2,\epsilon}}
      \!\!\! (S^{ab}T_a)_{;b}\, dV
   &=\int_{S^{1,\epsilon}_\tau\cup S^{2,\epsilon}_\tau}
   \!\!\! S^{ab} T_a n_b  \,dS
   - \int_{S^{1,\epsilon}_0\cup S^{2,\epsilon}_0}\!\!\! S^{ab} T_a n_b \,dS\\
   &\phantom{=}\quad
   + \int_{\Lambda^{2,\epsilon}_\tau}\!\!\! S^{ab} T_a \lambda_b \,d\Lambda
   - \int_{\Lambda^{1,\epsilon}_\tau}\!\!\! S^{ab} T_a \lambda_b \,d\Lambda
\endaligned
$$
The integrand on the left hand side may be written as
$$ (S^{ab}T_b)_{;a} = T^a\phi_{,a}(f-\phi) +
T^{c;d}\phi_{,c}\phi_{,d} - \tfrac12 T^a{}_{;a} (g^{cd}\phi_c\phi_d+\phi^2)
$$ 
Now $\phi\in H^1(\Omega)$ implies that
$$ \lim_{\epsilon\to0} 
   \int_{\Omega_\tau^{1,\epsilon}\cup\Omega_\tau^{2,\epsilon}}
   \!\!\! (S^{ab}T_a)_{;b}\, =
   \int_{\Omega_\tau} T^a\phi_{,a}(f-\phi) \,dV
   +\int_{\Omega_\tau} \bigl(T^{c;d}\phi_{,c}\phi_{,d} - \tfrac12
   T^a{}_{;a} (g^{cd}\phi_c\phi_d+\phi^2) \bigr)\,dV $$
and similarly $\phi_{|S_\tau}\in H^1(S_\tau)$ implies
$$ \lim_{\epsilon\to0}
   \int_{S^{1,\epsilon}_\tau\cup S^{2,\epsilon}_\tau}
   \!\!\! S^{ab} T_a n_b \,dS = E(\tau) $$
Finally we define the energy flux across $\Lambda$, which will exist for
$\phi\in H^1(\Omega)$;
$$ F(\tau)= \lim_{\epsilon\to0} \biggl(
   \int_{\Lambda^{2,\epsilon}_\tau}\!\!\! S^{ab} T_a n_b \,dS
   - \int_{\Lambda^{1,\epsilon}_\tau}\!\!\! S^{ab} T_a n_b \,dS
   \biggr) $$

The energy equation, in the limit is therefore
$$ E(\tau) = E(0) - F(\tau)
   + \int_{\Omega_\tau} \phi_{,t}(f-\phi) \,dV
   +\int_{\Omega_\tau} \bigl(T^{c;d}\phi_{,c}\phi_{,d} - \tfrac12
      T^a{}_{;a} (g^{cd}\phi_c\phi_d+\phi^2) \bigr)\,dV $$
The fact that $g_{ab}\in C^{1-}(\Omega)$, $T^a{}_{;b}\in L^\infty(\Omega)$
and $\phi\in H^1(\Omega)$ implies that for all $0<\tau<b$, there exist
constants $C_1,\,C_2>0$ such that
$$ E(\tau) \leq E(0) + \bigl|F(\tau)\bigr|
   + C_1 \int_{\Omega_\tau} |\phi_{,t}| \bigl( |f| + |\phi|
   \bigr) \,d^4x
   + C_2 \int_{\Omega_\tau} \Bigr( \sum_{a,\,b=0}^4 |\phi_{,a}||\phi_{,b}|
   + |\phi|^2 \Bigr) \,d^4x $$
On applying the Cauchy-Schwarz inequality
$$ E(\tau) \leq E(0) + \bigl|F(\tau)\bigr|
   + \tfrac12 C_1 \sobnormsq{0}{\Omega_\tau}{f}
   + (C_1+4C_2) \sobnormsq{1}{\Omega_\tau}{\phi} $$
By applying~\eqref{sandwich}
$$ E(\tau) \leq E(0) + \bigl|F(\tau)\bigr| + \tfrac12 C_1
\sobnormsq{0}{\Omega_\tau}{f} + {1\over B_1}(C_1+4C_2)
\int_{\tau'=0}^\tau E(\tau')\,d\tau' $$
Finally Gronwall's inequality gives
$$ E(\tau) \leq \bigl( E(0) + \bigl|F(\tau)\bigr|
   + \tfrac12 C_1 \sobnormsq{0}{\Omega_\tau}{f} \bigr)
   \exp \Bigl({C_1+4C_2\over B_1} \tau\Bigr)
   \tag\tagnum\xlabel{energy}$$
\qed\enddemo

Equivalently~\eqref{energy} may be written in terms of the Sobolev norms as
$$ \sobnormsq{1}{S_\tau}{\phi} \leq {1\over B_1} \bigl( B_2
   \sobnormsq{1}{S_0}{\phi} + \bigl|F(\tau)\bigr| + \tfrac12 C_1
   \sobnormsq{0}{\Omega_\tau}{f} \bigr) \exp \Bigl({C_1+4C_2\over
   B_1} b\Bigr) \tag\tagnum\xlabel{energy2}$$
This differs from the standard energy inequality by the presence of an
additional flux term $F(\tau)$. However if $F(\tau)$ vanishes then we are
able to establish that a solution is unique.

\proclaim{Proposition \procnum}
Solutions with a vanishing flux $\flux\tau\phi$ are unique.
\endproclaim

\demo{Proof}
Suppose that $\gamma$ is the difference of two such solutions of
\eqref{ivp}, then it must be a solution of the initial value problem
$$ \aligned
   \square\gamma & =0 \\
   \gamma_{|S} &= 0 \\
   n^a {\gamma_{,a}}_{|S} &= 0
   \endaligned
$$
and so the corresponding energy inequality is
$$ \sobnormsq{1}{S_\tau}{\gamma} \leq {1\over B_1} 
   \exp \Bigl({C_1+4C_2\over B_1} b\Bigr) |F(\tau)|
$$
Therefore it follows that if $\flux\tau\gamma=0$ then $\gamma=0$.
\qed\enddemo

A sufficient, but not necessary condition for $F(\tau)$ to vanish is that
the solution $\phi$ is $C^1$.

\subhead 2.2 Existence of solutions in $H^1(\Omega)$ \endsubhead

In order to prove that an $H^1$ solution $\phi$ to~\eqref{ivp} exists, we
shall prove the existence of $\psi\in H^1(\Omega)$ to the following initial
value problem with zero initial data
$$ \aligned
   \square \psi & =f-\square q \\
   \psi_{|S} &= 0 \\
   n^a {\psi_{,a}}_{|S} &= 0
   \endaligned
   \tag\tagnum\xlabel{ivp2}
$$
where
$$ q(t,x^\alpha) = \phi_0(x^\alpha) + t \phi_1(x^\alpha) $$
The choice of $q$ ensures that it satisfies the conditions
$$ \aligned
   q_{|S} &=\phi_0 \\
   n^a q_{,a|S} &=\phi_1
\endaligned
$$
so that $\phi=\psi+q$ will be a solution of~\eqref{ivp}.

It will be assumed that $\phi_0$ and $\phi_1$ are at least in $H^1(S)$,
which guarantees that $q\in H^1(\Omega)$; however the energy
inequality~\eqref{energy} will only be applicable to~\eqref{ivp2} if the
source term $f-\square q$ of~\eqref{ivp2} is also in $H^1(\Omega)$.  As the
connection is essentially bounded, sufficient conditions to achieve $\square
q\in H^1(\Omega)$ are $\phi_0,\,\phi_1\in H^3(S)$.

Our construction of a solution $\psi\in H^1(\Omega)$ to~\eqref{ivp2} will
follow the methods of Egorov and Shubin~(1992) and Clarke~(1998). We shall
work exclusively in the space $H^1(\Omega)$ as a Hilbert space with the inner
product
$$ \ip\psi\omega = \int_{\Omega} (e^{ab} \psi_{;a} \omega_{;b} +
   \psi\omega)\, dV $$
where $e^{ab}=g^{ab}+2T^a T^b$ is a positive definite metric. It should be
noted that in general we do not have $e^{ab}{}_{;c} =0$.

The induced norm is equivalent to the
Sobolev norm $\sobnorm{1}{\Omega}{\psi}$ in that there exist constants
$B_3,\,B_4>0$ such that for all $\psi,\,\omega\in H^1(\Omega)$
$$ B_3 \sobnormsq{1}{\Omega}{\psi} \leq \ip{\psi}{\psi} \leq B_4
   \sobnormsq{1}{\Omega}{\psi} \tag\tagnum\xlabel{ipnorm}$$

We shall define two subspaces of  $H^1(\Omega)$
$$ \aligned
   V_0 &= \{\, \psi\in H^3(\Omega)\cap C^2(\Omega) \,|\, 
    \psi_{|S}=n^a{\psi_{,a}}_{|S}
   =\psi_{,b|S}=n^a{\psi_{,ab}}_{|S}=0 \,\} \\
   V_1 &= \{\, \omega\in H^3(\Omega)\cap C^2(\Omega) \,|\, 
    \omega_{|S_{\tau_2}}=n^a{\omega_{,a}}_{|S_{\tau_2}}
   =\omega_{,b|S}=n^a{\omega_{,ab}}_{|S}=0 \,\} \\
\endaligned
$$

The condition that $V_0,\,V_1\subset H^3(\Omega)$ ensures that $\square
V_0,\,\square V_1\subset H^1(\Omega)$ whereas the $C^2$ differentiability
is sufficient to force the flux term for both $\psi$ and $\psi_{;a}$ in the
energy inequality~\eqref{energy} to vanish. The energy inequality for
$\psi\in V_0$ then becomes
$$ \sobnormsq{1}{S_\tau}{\psi}
   \leq {A\over 2B_1} e^{A\tau_2/B_1}
   \sobnormsq{0}{\Omega_\tau}{\square\psi}  $$
In particular we apply this to the region $\Omega^+$, integrate over the
time variable $\tau$ and apply~\eqref{ipnorm} to give a bound for
$\ip{\psi}{\psi}$
$$ \ip\psi\psi \leq C_3 \sobnormsq{1}{\Omega}{\square\psi} 
   \quad \forall\psi\in V_0
   \qquad\text{$C_3>0$ constant} $$
We may obtain a similar inequality for $\omega\in V_1$, by regarding
$S_{\tau_2}$ as an initial surface, with zero initial data, evolving back
in time and constructing an analogous energy inequality; thus we obtain
$$ \ip\omega\omega \leq  C_4 \sobnormsq{1}{\Omega}{\square\omega} \quad
   \forall\omega\in V_1 
   \qquad\text{$C_4>0$ constant}
   \tag\tagnum\xlabel{boxomega} $$

The construction of $V_0$, $V_1$ and $\ip\psi\omega$ are motivated by the
need to have $\square$ behaving as a self-adjoint operator, however in
general this cannot be guaranteed and extra conditions on the space-time
are needed to achieve this.

\proclaim{Lemma \procnum}
\xlabel{selfadj}
If the following conditions are satisfied
\roster
\item $T^a{}_{;b}=0$
\item $R_{ab} T^b=0$
\endroster
then
$$ \ip{\square\psi}{\omega}=\ip{\psi}{\square\omega}
\qquad\forall \psi\in V_0,\, \omega\in V_1 $$
\endproclaim

\demo{Proof}%
Let $\psi\in V_0$ and $\omega\in V_1$.
We apply Stokes' theorem to the region
$\Omega^{1,\epsilon}_b\cup\Omega^{2,\epsilon}_b$ and take the limit
$\epsilon\to 0$
$$\aligned
\int_{\Omega^+} ( \psi^{;c}\omega_{;c} + (\square\psi)\omega ) \,dV
&= \int_{S_b} \psi_{;c} \omega n^c \,dS
- \int_{S} \psi_{;c} \omega n^c \,dS\\
&\qquad+ \int_{\Lambda^2} \psi_{;c} \omega \lambda^c \,d\Lambda
- \int_{\Lambda^1} \psi_{;c} \omega \lambda^c \,d\Lambda \\
\int_{\Omega^+} ( \psi^{;c}\omega_{;c} + \psi\square\omega ) \,dV
&= \int_{S_b} \psi \omega_{;c} n^c \,dS
- \int_{S} \psi \omega_{;c} n^c \,dS\\
&\qquad+ \int_{\Lambda^2} \psi \omega_{;c} \lambda^c \,d\Lambda
- \int_{\Lambda^1} \psi \omega_{;c} \lambda^c \,d\Lambda \\
\int_{\Omega^+} \bigl( e^{ab} \psi_{;a}{}^{;c} + e^{ab;c} \psi_{;a}
\omega_{;bc} + e^{ab} \psi_{;a} \square(\omega_{;b}) \bigr) \,dV
&= \int_{S_b} e^{ab} \psi_{;a} \omega_{;bc} n^c \,dS
- \int_{S} e^{ab} \psi_{;a} \omega_{;bc} n^c \,dS\\
&\qquad + \int_{\Lambda^2} e^{ab} \psi_{;a} \omega_{;bc} \lambda^c \,d\Lambda
- \int_{\Lambda^1} \psi_{;a} \omega_{;bc} \lambda^c \,d\Lambda \\
\int_{\Omega^+} \bigl( e^{ab} \psi_{;a}{}^{;c} + e^{ab;c} \psi_{;a}
\omega_{;bc} + e^{ab} \psi_{;a} \square(\omega_{;b}) \bigr) \,dV
&= \int_{S_b} e^{ab} \psi_{;a} \omega_{;bc} n^c \,dS
- \int_{S} e^{ab} \psi_{;a} \omega_{;bc} n^c \,dS\\
&\qquad + \int_{\Lambda^2} e^{ab} \psi_{;a} \omega_{;bc} \lambda^c \,d\Lambda
- \int_{\Lambda^1} \psi_{;a} \omega_{;bc} \lambda^c \,d\Lambda \\
\endaligned
$$
It will be noted that the contributions from the surfaces $S_0$ and $S_b$
vanish.  With $\psi,\,\omega\in C^2$ the flux terms will also vanish.

On combining the above equations
$$
\multline
\int_{\Omega^+} \bigl( (\square\psi)\omega + e^{ab} \square(\psi_{;a})
\omega_{;b} \bigr) \,dV + \int_{\Omega^+} e^{ab;c} \psi_{;ac} \omega_{;b}
\,dV \\ 
= \int_{\Omega^+} \bigl( \psi\square\omega + e^{ab} \psi_{;a}
\square(\omega_{;b}) \bigr) \,dV + \int_{\Omega^+} e^{ab;c} \psi_{;a}
\omega_{;bc} \,dV 
\endmultline
$$
and on applying the contracted Ricci identity
$$ \square(\psi_{;a}) = (\square\psi)_{;a} + R^b{}_a \psi_{;b} $$
we obtain
$$ \split
\int_{\Omega^+} \bigl( (\square\psi)\omega + e^{ab} (\square\psi)_{;a}
\omega_{;b} \bigr) \,dV &=
\int_{\Omega^+} \bigl( \psi\square\omega + e^{ab} \psi_{;a}
(\square\omega)_{;b}\bigr) \,dV \\
&+ 2\int_{\Omega^+} T^a R^{bc}T_c (\psi_{;a}\omega_{;b} -
\psi_{;b}\omega_{;a} )\,dV\\
&+ \int_{\Omega^+} e^{ab;c} (\psi_{;a} \omega_{;bc} - \psi_{;ac}
\omega_{;b}) \,dV  
\endsplit
$$
Using the fact that $e^{ab}=g^{ab}+2T^aT^b$ and applying the hypotheses to
this equation, it is now evident that
$$ \ip{\square\psi}{\omega}=\ip{\psi}{\square\omega}.$$
\qed\enddemo

As a consequence we may define a linear functional
$k:\square V_1\to\Real$ by 
$$ k(\square\omega)=\ip{f-\square q}{\omega} $$

\proclaim{Proposition \procnum}
$k$ defines an function $\psi\in H^1(\Omega^+)$ with
$\psi$ and $\psi_{,t}$ vanishing on $S$.
\endproclaim

\demo{Proof}
This result is a consequence of the Hahn-Banach theorem, in conjunction
with the Riesz representation theorem. In order to apply these theorems, we
must show that
\roster
\item $k$ is bounded by a semi-norm on $\square V_1$
\item $\square V_1$ is a dense subspace of $H^1(\Omega^+)$
\endroster

A bound for $k$ may be constructed by applying~\eqref{boxomega};
$$ \xalignat2
    \left|k(\square\omega)\right|
    &= | \ip{f-\square q}{\omega} | \\
    & \leq \ip{f-\square q}{f-\square q}^{1/2} \ip\omega\omega^{1/2} \\
    & \leq {C_2}^{1/2} {B_4}^{1/2} \sobnorm{1}{\Omega^+}{f-\square q}
    \sobnorm{1}{\Omega^+}{\square\omega}
\endxalignat $$

To prove that $\square V_1$ is dense in $H^1(\Omega^+)$, It is sufficient
to show that any function in $(\square V_1)^\perp$ is necessarily zero.
Suppose that $\lambda\in (\square V_1)^\perp$ then, because $V_0$ is dense,
there exists a sequence $(\psi_n)$ in $V_0$ converging to $\lambda$. By
using Lemma~\xref{selfadj} and taking the limit $n\to\infty$, we have for
all $\omega\in V_1$
$$ \ip{\square\psi_n}{\omega} = \ip{\psi_n}{\square\omega}
   \to \ip{\lambda}{\square\omega} = 0$$
Since $V_1$ is dense, this implies that $\square\psi_n\to0$ almost
everywhere and, by~\eqref{boxomega}, we have $\psi_n\to0$ almost everywhere.
Therefore $\lambda=0$ in $H^1(\Omega^+)$.
\qed
\enddemo

We have therefore proved the following

\proclaim{Theorem \procnum\xlabel{final}}

Let $(M,g)$ be a space-time containing a singular hypersurface $\Lambda$
with $g_{ab}$ being $C^{1-}$ across $\Lambda$. Moreover suppose there
exists a congruence of time-like geodesics with a tangent field $T^a$
satisfying
\roster
\item $T^{a}_{;b}=0$
\item $R_{ab} T^b =0$
\endroster
then solution $\phi\in H^1(\Omega^+)$ exists to the initial value problem
$$ \aligned
   \square \phi & =f \\
   \phi_{|S} &= \phi_0 \\
   n^a {\phi_{,a}}_{|S} &= \phi_1
   \endaligned
$$
where
$$ \gathered
    f\in H^1(\Omega^+) \\
    \phi_0,\,\phi_1\in H^3(S) \\
   \endgathered $$
Moreover if a solution $\phi$ satisfies the vanishing flux condition
$\flux\tau\phi=0$, then it is unique.
\endproclaim

If we replace conditions (1) and (2) by the weaker condition that
$T^a{}_{;b}$ is essentially bounded then $\square$ is only a self adjoint
operator in $H^0(\Omega)$ with the usual $L^2$ inner product. This is
sufficient to prove existence of solutions in $H^0(\Omega)$.

\head 3. Examples \endhead

\subhead Thin shells of matter \endsubhead

An elementary example of a space-time containing a hypersurface singularity
may be constructed using the following metric
$$  ds^2= -dt^2+ F(z)^2 dx^2 + G(z)^2 dy^2 +dz^2 $$
where
$$ \aligned
   F(z) &= 1-azH(z) \\
   G(z) &= 1+ az\bigl(1-H(z)\bigr)
\endaligned $$
It may be shown that the non zero components of the curvature tensor are
$$ \aligned
   R_{xzxz} &= a \delta(z) \\
   R_{yzyz} &= a \delta(z)
\endaligned $$
and the non-zero components of the energy-momentum tensor are
$$ \aligned
   T_{tt} &= 2a\delta(z) \\
   T_{xx} &= a\delta(z) \\
   T_{yy} &= a\delta(z)
\endaligned$$
and with $a>0$ this space-time represents a positive energy matter
distribution whose support is the hypersurface $z=0$.

This space-time possesses a metric which is $C^{1-}$ across the singular
hypersurface and moreover the integral curves of $\partial/\partial t$
forms a congruence of time-like geodesics whose tangent satisfies the
conditions $T^a{}_{;b}=0$ and $R_{ab} T^b=0$.
Therefore Theorem~\xref{final} is applicable and we are able to establish
that hyperbolicity is applicable

\subhead Impulsive gravitational waves \endsubhead

Another example of a space-time with a singular hypersurface that satisfies
similar differentiability conditions is one which contains impulsive
gravitational waves. In double null coordinates
$$ u=\tfrac{1}{\surd2} (t-z),\quad v=\tfrac{1}{\surd2}(t+z)$$
the metric may be written as
$$  ds^2= -2du\,dv + F(u)^2 dx^2 + G(u)^2 dy^2 $$
where
$$ \aligned
   F(z) &= 1+auH(u) \\
   G(z) &= 1-auH(u)
\endaligned $$
The space-time is vacuum, but has a Weyl tensor with delta function
components
$$ \aligned
   C_{uxux} &= -a\delta(u) \\
   C_{uyuy} &=  a\delta(u)
\endaligned $$

A congruence of time-like geodesics whose tangent field is
$$T^a=(\tfrac{1}{\surd2},\tfrac{1}{\surd2},0,0)$$
may be constructed; however $T^a_{;b}$ is essentially bounded but not
vanishing, so theorem~\xref{final} is not applicable to this space-time in
its current form.

The proof of Theorem~\xref{final} breaks down at Lemma~\xref{selfadj} in
that self-adjointness of $\square$ cannot be established in $H^1$. At most
we are able to show existence in $H^0$ by defining the subspaces $V_0$ and
$V_1$ to be in $H^0$ and work with the $H^0$ inner product, with which
$\square$ will be self-adjoint without the need for $T^a{}_{;b}$ to vanish
completely. 

\head 4. Conclusion \endhead

Theorem~\xref{final} gives us sufficient conditions on the initial data for
the existence of an $H^1$ solution in space-times with a singular
hypersurface; provided certain conditions on the existence of a certain
time-like geodesic congruence have been met.

It was found that stronger conditions on the initial data are required,
than what is required in Mink\-owsk\-i space. However the fact that the 
connection components were essentially bounded avoided the need for
imposing extra conditions involving the rate at which spatial derivatives
vanished as were needed for the conical case (Wilson 2000).

The behaviour of $T^a{}_{;b}$ was found to be more of an issue here as
although $T^a{}_{;b}$ being essentially bounded is sufficient for
establishing energy inequalities, it was found that the vanishing of
$T^a{}_{;b}$ was required $\square$ to be self adjoint in  $H^1$ and
therefore to enable existence to be proved in this space. The conical
space-time (Wilson, 2000) has vanishing $T^a{}_{;b}$

\enddocument

\Refs

\ref
\by R.~Abraham, J.~E~Marsden and T.~Ratiu
\book Manifolds, Tensor Analysis, and Applications
\bookinfo Springer Applied mathematical Sciences
\publ Springer
\vol 75
\yr 1988
\endref

\ref
\by H.~Balasin and H.~Nachbagauer
\paper What curves the Schwarzschild Geometry
\jour Classical and Quantum Gravity
\vol 10
\pages 2271--78
\yr 1993
\endref

\ref
\by Y.~Choquet-Bruhat, C.~De Witt-Morette and M.~Dillard-Bleick
\book Analysis, Manifolds and Physics
\publ North-Holland
\yr 1977
\endref

\ref 
\by C.~J.~S.~Clarke
\paper Singularities: boundaries or internal points ?
\inbook Singularities, Black Holes and Cosmic Censorship
\ed P.~S.~Joshi
\publ IUCAA, Bombay
\pages 24--32
\yr 1996
\endref

\ref
\by C.~J.~S.~Clarke
\paper Generalised hyperbolicity in singular space-times
\jour Classical and Quantum Gravity
\vol 15
\pages 975--84
\yr 1998
\endref

\ref
\by C.~J.~S.~Clarke and N.~O'Donnell
\paper Dynamical extension through a space-time singularity
\jour Rendiconti del Seminario Matematico, Universit\'a Torino
\vol 50
\pages 39--60
\yr 1992
\endref

\ref
\by C.~J.~S.~Clarke, J.~A.~Vickers and J.~P.~Wilson
\paper Generalised functions and distributional curvature of cosmic
  strings
\jour Classical and Quantum Gravity
\vol 13
\pages 2485--98
\yr 1996
\endref

\ref
\by J.~F.~Colombeau
\book New generalised functions and multiplication of distributions
\bookinfo  North-Holland Mathematics Studies 84
\publ North-Holland
\yr 1984
\endref

\ref
\by Y.~V.~Egorov and M.~A.~Shubin
\book Partial differential Equations
\vol 1
\yr 1992
\publ Springer
\endref

\ref
\by M.~Grosser, M.~Kunzinger, R.~Steinbauer and J.~Vickers
\paper A global theory of algebras of generalised functions
\jour Preprint. {\it math/9912216}
\yr 1999
\endref

\ref
\by S.~W.~Hawking and G.~F.~R.~Ellis
\book The large scale structure of space-time
\publ Cambridge University Press
\yr 1973
\endref

\ref
\by R.~Penrose
\paper The geometry of impulsive gravitational waves
\inbook General relativity: papers in honour of J.~L.~Synge
\ed L.~O'Raif\-eartaigh
\publ Clarendon Press, Oxford
\yr 1972
\endref

\ref
\by J.~A.~Vickers
\paper Generalised cosmic strings
\jour Classical and Quantum Gravity
\vol 4
\pages 1--9
\yr 1987
\endref

\ref
\by J.~A.~Vickers and J.~P.~Wilson
\paper Generalised hyperbolicity in conical space-times
\jour Classical and Quantum Gravity
\vol 17
\pages 1333--60
\yr 2000
\endref

\endRefs
\bye